\documentclass[aps,prc,showpacs,twocolumn]{revtex4}
\usepackage{graphicx}
\usepackage{dcolumn}
\usepackage{bm}
\usepackage{longtable}
\usepackage{hyperref}
\usepackage{natbib}
\usepackage{amsmath}
\begin{document}
\title{Investigating $\beta$-decay properties of  spherical nuclei along the
possible r-process path}
\author{Dong-Liang Fang$^{a,b}$, B. Alex Brown$^{a,b,c}$ and Toshio~Suzuki$^{d,e}$}
\affiliation{$^a$National Superconducting Cyclotron Laboratory, Michigan State
University, East Lansing, Michigan 48824, USA}
\affiliation{$^b$Joint Institute for Nuclear and Astrophysics, Michigan State
University}
\affiliation{$^c$Department of Physics and Astronomy, Michigan State University,
East Lansing, MI 48824, USA}
\affiliation{$^d$Department of Physics, College of Humanities and Sciences,
Nihon University, Sakurajosui 3-25-40, Setagaya-ku, Tokyo 156-8550, Japan}
\affiliation{$^e$National Astronomical Observatory of Japan,
Mitaka, Tokyo 181-8588, Japan}
\begin{abstract}
The spherical QRPA method is used for
the calculations of the $\beta$-decay properties of the neutron-rich nuclei in
the region
near the neutron magic numbers N=82 and  N=126 which are important for
determination of the
r-process path. Our calculations differ from previous works by the
use of realistic forces for the proton-neutron interaction.
Both the allowed and first-forbidden $\beta$-decays are
included. Detailed comparisons with the experimental measurements and the
previous
shell-model calculations are performed. The results for half-lives and
beta-delayed neutron
emission probabilities will serve as input for the r-process nucleosynthesis
simulations.
\end{abstract}
\maketitle

\section{Introduction}
The synthesis of the heavy elements in the universe is one of the important and
interesting topics in modern physics. Different processes and astrophysical
events are involved in understanding the measured isotopic abundances. One
kind of process is the so-called rapid-neutron capture process (r-process),
that gives rise to heavy elements beyond iron in our solar
system~\cite{WHW02,CTT91}. The site of this process is still not clear. One
of the
popular ideas is the high-entropy wind of core collapse Type II
supernovae~\cite{CTT91,FKMPJTT09,FKPRTT10}. The obstacles to the simulation of
the
r-process are two-fold. (1) The unclear site gives uncertainties to the
astrophysical environmental parameters which are crucial for the
initial conditions of
the simulations of the r-process evolution, such as the initial neutron
richness, the temperature {\it etc}~\cite{FKPRTT10}. (2) Most of
the nuclei involved in the evolution are exotic neutron-rich ones which are
currently out of the reach of experiments, so their reaction
and decay rates are still uncertain.
For the r-process the nuclei are very neutron-rich and some are near the neutron
dripline.
The general pattern of solar element abundance shows
peaked distributions around $A\sim130$ and $A\sim 190$.
Surveys~\cite{CTT91} show that in order to produce these two heavy element
peaks,
the properties of nuclei around $A\sim 130$ and $190$ are important especially
those near the proton or neutron magic numbers.

The nuclear chart can be divided into regions near the magic numbers
where the low-lying
states are best described in a spherical basis, and regions in between
the magic numbers where they are best described in a deformed basis.
The QRPA method has been developed for each of these. There is
a transitional region that, at present, must be interpolated in terms
of the spherical and deformed limits.
In a previous paper we focused on the deformed region of nuclei centered
on Z=46 and N=66~\cite{FBS12}.
In this
work we will focus on the heavy spherical nuclei near the magic numbers. In
these
regions, some shell-model calculations can be
performed with a truncated model space. Early calculations only included
Gamow-Teller (GT) decay~\cite{ML99,LM03}, but recent calculations
~\cite{SYKO12,CMLNB07,ZCCLMS13} have also included
First-Forbidden (FF) decays which turn out to be important for
the $N=126$ isotones~\cite{SYKO12}. QRPA methods have previously been 
used,
such as the QRPA method with separable forces in~\cite{MR90,MPK03,HBHMKO96}, the
self-consistent QRPA from DFT~\cite{EBDBS99} and the continuum
QRPA~\cite{BG00,Bor03}
{\it etc}.
As an improved alternative, we propose the pn-QRPA method with realistic forces
as introduced in~\cite{STF88,CFT87}. The advantage of this method is that with
the
G-matrix obtained from the Bethe equations with realistic potentials fitted from
the nucleon-nucleon scattering data, one can obtain the full spectrum for the
ground as well as excitation states of the odd-odd daughter nuclei. The full
spectrum provides an exact treatment for the excitation energies which are
missing in most other QRPA methods. The inclusion of states with more
spin-parities also means that we can deal with the negative parity FF
transitions which are missing in some calculations~\cite{MR90,MPK03}.

This article is arranged as follows. In next section the
QRPA method and its application to allowed and forbidden beta decay is outlined.
The choice of model parameters is discussed in Sec. III.
The results are presented in Sec. IV in comparison with experiment
and to previous calculations. Sec. V presents the conclusions.

\section{Formalism and Method}
In this work, we will calculate both the allowed Gamow-Teller (GT) and first
forbidden (FF) $\beta$-decays for proton-neutron even-even and odd-even nuclear
isotones near $N=82$ and $N=126$.
The half-lives for these decays can be expressed generally
as:
\begin{eqnarray}
 T_{1/2}=\frac{\ln 2}{\Gamma}.
\end{eqnarray}
Here $\Gamma=\sum_i\Gamma_i$ is the total decay width and is a sum over all the
possible decay widths from the ground state of initial parent nucleus to
different states of final daughter nucleus with the specific selection rules.
For GT decay, the width can be expressed as:
\begin{eqnarray}
 \Gamma_i^{{\rm GT}}=({\it f_0}/K_0)g_A^2 B_i({\rm GT}^-).
\label{GT}
\end{eqnarray}
The term ${\it f_0}$ is the dimensionless phase-space factor depending on
$\beta$-decay Q value~\cite{BB82}, while $K_0$ is a
combinations of constants defined as $K_0=\frac{2 \pi^3 \hbar^7}{m_e^5 c^4}$.
The GT strength $B({\rm GT}^-)$ can be expressed in terms of the reduced matrix
element for spherical nuclei $B({\rm GT}^-)=|M_i({\rm GT}^-)|^2/(2 J+1)$, here
$J$ is the
spin of the parent nucleus. A frequently used quantities here is the log$ft$
value, which is
defined here as log$ft$=log$[\frac{C}{g_A^2 B({\rm GT}^-)}]$ with $C=\ln 2 K_0=
6170$.

For FF decay, the expression is more complicated, as in ~\cite{BB82}:
\begin{eqnarray}
\Gamma_i^{{\rm FF}}=\frac{f_i}{8896} (s^{-1}),
\end{eqnarray}
with
\begin{eqnarray}
f_i=\int_{1}^{\omega} \mathcal{C}(\omega)F(Z,\omega)p \omega (\omega_0-\omega)^2
d\omega,
\end{eqnarray}
and with $\mathcal{C}(\omega)$ defined as
\begin{eqnarray}
\mathcal{C}(\omega)=k+ka\omega+kb/\omega+kc \omega^2.
\label{FF}
\end{eqnarray}
$\omega$ is the electron energy in the unit of electron mass
$\omega=E_e/m_ec^2$, $F(Z,\omega)$ is the Fermi function as expressed
in~\cite{BB82}
and $k$, $ka$, $kb$ and $kc$ are the nuclear matrix elements depending on the
nuclear structure. The detailed expression of these matrix elements are given in
eq.(8) in Ref.~\cite{ZCCLMS13} and eq.(3,4,5)
in Ref.~\cite{SYKO12}. The log$ft$ values in this case are defined as
log$ft$=log($f_0C/\Gamma_i^{{\rm FF}}$), here $f_0$ is the phase space factor
for GT decay.

In this work, for the matrix element calculations, we adopt the pn-QRPA
(proton-neutron Quasi-particle Random Phase Approximation) method. The concept
of quasi-particle starts with the nuclear pairing. The most common description
of
pairing in nuclear physics is the BCS formalism. Under the BCS formalism, one
can define the quasi-particle operator $\alpha_\tau=u_\tau c_\tau+v_\tau
\tilde{c}_\tau^\dagger$, where $u_\tau$, $v_\tau$ are the BCS coefficients
solved
from BCS equations, and $c^\dagger_{\tau}$ is the single particle creation
operator. With the quasi-particle operators, one can define the QRPA phonons
as~\cite{STF88}:
\begin{equation}
Q^{m\dagger}_{J^{\pi}M}=\sum_{pn}
{X^{m}_{pn}A^\dagger_{pn,J^{\pi}M}-Y^{m}_{pn}\tilde{A}_{pn,J^{\pi}M}}.
\end{equation}
Here the two quasi-particle operators are defined as
$A^\dagger_{pn,{J^{\pi}M}}\equiv [\alpha_{p}^\dagger\alpha_{n}^\dagger]_{J^{\pi}M}$, with $p$,
$n$ being proton and neutron respectively. The coefficients $X$'s and $Y$'s here
are the forward and backward amplitudes respectively, they can be derived from
the solutions of QRPA-equations~\cite{CFT87,STF88}:
\begin{eqnarray}
\left( \begin{array}{cc}
A & B  \\
B & A  \end{array} \right)
\left(\begin{array}{c}X \\ Y\end{array}\right)
=\omega\left(\begin{array}{cc}1 & 0 \\ 0 & -1 \end{array}\right)
\left(\begin{array}{c}X \\ Y\end{array}\right).
\end{eqnarray}
Here, $A_{pn,p'n'}=[A_{pn},[H,A^\dagger_{p'n'}]]$ and
$B_{pn,p'n'}=[{A}_{pn}^\dagger,[H,\tilde{A}_{p'n'}^\dagger]]$.
The Hamiltonian and the detailed expressions of $A$ and
$B$ with realistic interactions were presented in Ref.~\cite{STF88}.
In this scenario, the different excited states are defined as:
$|J^{\pi}M;m>\equiv Q^{m\dagger}_{J^{\pi}M}|0>$ with the QRPA phonon
$Q^{m}_{J^{\pi}M}$ acted on even-even vacuum $|0>$ .

For decays of even-even nuclei
we choose the BCS vacuum as
the ground state:
\begin{eqnarray}
|0\rangle_i=|QRPA\rangle\approx|BCS\rangle.
\end{eqnarray}
While the final states in the  odd-odd nuclei are the pn-QRPA excited
states as defined by:
\begin{eqnarray}
|m\rangle_f=Q^{m}_{J^\pi}|0\rangle.
\end{eqnarray}
We choose the state with the smallest eigenvalue from QRPA equations to be the
ground state of odd-odd final nucleus. The energies of all the excited states
are then:
$E_m=\omega_{m,J^\pi}-\omega_{g.s.}$, here the $\omega_m$ refer to the
eigenvalues
of the $m$th state from the QRPA solutions in the spherical systems while $\omega_{g.s.}$ is the smallest eigenvalues of the solutions. 
The
effective Q values for each state are $Q_m=Q-E_m$, where Q is the mass
difference
between the two ground states of parent and daughter nuclei.

To obtain the
lifetime, besides the Q values, one also needs the matrix elements in
equations \ref{GT}-\ref{FF}, for even-even to odd-odd decay.
These can be expressed in our formalism as:
\begin{eqnarray}
\langle J^\pi m||\tau^+ O_{I}^{K^\pi}||0\rangle_i=\delta_{J,K}\sum_{pn}\langle
p||\tau^+ O_{I}^{K\pi}||n\rangle \\ \nonumber
\times (X^{J^\pi,m}u_{p}v_{n}+Y^{J^\pi,m}v_{p}u_{n}).
\end{eqnarray}
Here the $O_{I}^{K^\pi}$ is the nuclear transition operator for $\beta$-decay.
The reduced matrix elements from the Wigner-Eckart theorem
are independent of M (the projection of total angular momentum on z axis). 
For the allowed decay, $O_{GT}^{1^+}={\bf \sigma}$ with the selection rule
$\Delta J=K=1$ and $\Delta \pi=1$, while for first-forbidden decay, the
expressions of the operators are more complicated (the detailed forms given
in ~\cite{SYKO12}) with the selection rules $\Delta J=K=0,~1,~2$ and
$\Delta \pi=-1$.

For decays of odd-even nuclei, one has an unpaired nucleon, the simplest
scenario is that given in Ref.~\cite{MR90}.
One quasi-particle or one quasi-particle plus one
QRPA-phonon acts on the BCS vacuum (the spectator mode in ~\cite{MR90}). In
Ref.~\cite{MR90}, for the single quasi-particle excitations, the correction from
the
coupling between the single quasi-particle and the QRPA-phonon was taken
into consideration with the assumption of the weak coupling limit. In our
calculation we find that this correction gives only small changes.
Since it takes a much longer time for the calculation, we neglect
this effect in the present calculation.

We have two kinds of states for
odd-even nuclei. First, the single particle state as given by:
\begin{eqnarray}
|\tau\rangle_i=\alpha_{\tau_i}^\dagger|0\rangle.  \nonumber
\end{eqnarray}
Here $\tau$ can be either proton or neutron. The ground states of the parent
nuclei are chosen to be the one with the lowest quasi-particle energies, and
this
also holds for the even-odd daughter nuclei. For all the other single
quasi-particle excitations, the relative excitation energies to the ground state
are $E_i=E_{\tau,i}-E_0$.

Another kind of states for the daughter nuclei is the quasi-particle plus phonon
state mentioned above without the possible mixing between then:
\begin{eqnarray}
|\omega_{K^\pi M',m},\tau';J^\pi M
\rangle=C_{K,M';j_{\tau'}m_{\tau'}}^{J^\pi,M}Q_{K^\pi,m}^\dagger\alpha_{\tau'}^\dagger|0\rangle.
\end{eqnarray}
The energies of such states are determined as follows. If we compare such states
with the odd-odd nuclei, the only
difference is the spectator single quasi particle, the difference of excitation energies in the two systems is
 simply equivalent to the difference of the Q values in these two systems (corresponding the difference of the ground states), this gives
$E_{m,0}=Q_{oe}-Q_{ee}+\omega_{m}-\omega_{g.s.}$.

The matrix elements for $\beta$-decay of odd-even nuclei to these two different
kinds of final states can be written as:
\begin{eqnarray}
\langle n_i||\tau^+O^{K^\pi}_{I}||p_0\rangle=u_{p_0} u_{n_i} \langle
p_0||\sigma||n_i\rangle    \nonumber
\end{eqnarray}
and
\begin{eqnarray}
\langle \omega_{K^\pi,m},p';J^\pi
M||\tau^+O^{K^\pi}_{I}||p_0\rangle=-\sqrt{(2J+1)(2j_{p_0}+1)}&&  \\ \nonumber
\times \delta_{p_0,p'} \left\{
\begin{array}{ccc}
K & J & j_{p_0} \\
j_{p_0} & 0 & K
\end{array}
\right\} \langle \omega_{K^\pi,m}||\tau^+ O^{K^\pi}_{I}||0\rangle&&
\end{eqnarray}
With $O_I^{K\pi}$ the same form as for the even-even case.

With these derived excitation energies and matrix elements, we can calculate the
beta decay properties with the results presented in the next sections.


\section{Choice of Parameters}

\begin{table}
\centering
\caption{The experimental $1^+$ energies and corresponding log$ft$ values of
the largest decay branch are listed here for different even-even Cd isotopes,
also the corresponding half-lives are shown. A comparison between the
experimental measurements and theoretical calculations has been made with
quenching factors $Q\equiv g_A/g_{A0}=0.4$. Here $g_{A0}=1.26$ is the
axial-vector-coupling constant for free neutrons.}
\begin{tabular}{|c|c|ccc|c|cc|cc|}
\hline
 & & \multicolumn{3}{c|}{Exp.~\cite{nucldata,Dill03}}& Calc.&
\multicolumn{2}{c|}{$Q=0.4$}\\ 
& N & $E_{1^+}$& log$ft$ &$t_{1/2}(s)$&$E_{1^+}$ & log$ft$ &$t_{1/2}(s)$\\
\hline
$^{118}$Cd &70 &0 &3.91&3018(12)&0&3.84&2484\\
$^{120}$Cd &72 &0 &4.10&50.80(21)&0&3.90&32.2\\ 
$^{122}$Cd &74 &0 &3.95&5.24(3)&0.04&3.97&4.76\\ 
$^{124}$Cd &76 & & &1.25(2)&0.40&4.02&1.79\\ 
$^{126}$Cd &78 & & &0.515(17)&0.70&4.09&0.67\\ 
$^{128}$Cd &80 &1.17&4.17&0.28(4)&1.02&4.11&0.25\\ 
$^{130}$Cd & 82 &2.12 &4.10 &0.162(7)&1.18&4.14&0.12\\ 
$^{132}$Cd &84 & & &0.097(10)&4.62&4.27&0.14\\ 
\hline
\end{tabular}
\label{cd}
\end{table}

In this work, we are interested in nuclei near or on the neutron magic numbers
82 and 126.
We chose the isotonic chains with
$N$=80, 82 and 84 and $N$=124, 126 and 128. Many of these nuclei are on
the r-process path and their decay properties are important
for r-process path that determines the final element productions near the peaks.

For single-particle (SP) energies, we adopt here those derived from solution of
Hartree-Fock equations with the SkX Skyrme interaction~\cite{SKX}.
For the unbound positive-energy states, we make an approximate extrapolation
for their discrete energies.
We choose the QRPA model space as follows. We include all the SP levels
with energies up to 5 MeV for neutrons and protons.
For the pairing part, the BCS equations are solved with constant
pairing gaps obtained from the symmetric five-term formula ~\cite{AW03}, where
the pairing gaps are derived from the odd and even mass differences. The obtained
BCS coefficients and quasi-particle energies are then used as inputs for the
QRPA equations as described above. For odd-even nuclei, the BCS solutions
are especially important for nuclei with small Q values, since they determined
which single-particle transitions are important for the lowest energies.


\begin{table}
\centering
\caption{Decay schemes for three even-even isotopes near $^{132}$Sn. 
For understanding the FF decays, we show the spin-parties, excitation
energies and log$ft$ values for several important FF decay branches. 
For the two theoretical log$ft$ values, 
the one without subscript uses the queching factors $g_A=0.5g_{A0}$ and $g_V=0.5g_{V0}$ while the 
subscript $u$ means no quenching has been taken into account. For both cases, an enhancement factor 
$\epsilon=2$ for the tensor part of $0^-$ transitions is adopted \cite{War91}.}
\begin{tabular}{|c|c|ccc|cccc|}
\hline
                   &                     &     & Exp.\cite{nucldata}& & & Theo.& &\\
                   &  $J^{\pi}_i$ & $J^{\pi}_f$ & $E_{ex}$ & log$ft$ &log$ft_q$ &log$ft_u$ &$J^{\pi}_f$ & $E_{ex}$ \\
\hline
                   &  $0^+$     & $1^-,0^-$   & 0.080    & 6.14     &  6.15       &    5.55     & $0^-$       &  0         \\
$^{140}$Xe &                 & $(0,1^-)$     & 0.515    & 6.82      & 6.58       &    5.98     & $1^-$       &  0.127   \\
                   &                & $0^{(-)},1^{(-)}$ & 0.653& 5.98     &  5.90     &     5.29     & $1^-$       &  0.586   \\
                   &                & $(1,2^-)$     & 0.800    &$\approx$7.1& 7.36&     6.22    & $2^-$       &  0.370    \\
                   &                & $1^{(-)}$      & 0.966    & 6.77      &  6.56      &     5.95     & $1^-$       &  1.350   \\
\hline
                   &  $0^+$& $2^-$             & 0.011    & $>$8.5  & 7.62       &    7.01       & $2^-$       & 0.041      \\
$^{138}$Xe &             & $(1)^-$          & 0.016    & 7.2        &  7.01      &    6.41       & $1^-$       & 0.111      \\
                   &             & $1^-,2^-$     & 0.258    & 7.32     &  7.42       &    6.82       & $2^-$       & 0.349      \\
                   &             & $1^-$            & 0.412    & 6.79     &  6.29       &    5.68       & $1^-$       & 0.563      \\
                   &             & $0^-,1^-$     & 0.450    & 6.59     &  6.52       &    5.92       & $0^-$       & 0             \\
\hline
                   & $0^+$& $(1^-)$          & 0            & $>$6.7 & 6.72        &    6.16       & $1^-$       & 0.169 \\
$^{136}$Te &            & $(0^-,1,2^-)$& 0.222     & 7.23     & 6.79        &    6.19       & $2^-$       & 0.540  \\
                   &            & $(0^-,1)$      & 0.334     & 6.27     &  6.00       &    5.39       &  $1^-$      & 0.749   \\
                   &            & $(0^-,1)$      & 0.631     & 6.28     &  6.37       &    5.77       &  $1^-$       & 0         \\
                   &            & $(0^-,1,2^-)$& 0.738    & 7.57     &  7.70       &    7.09       &  $2^-$      & 0.194    \\
\hline
\end{tabular}
\label{ff82ee}
\end{table}


\begin{table}
\centering
\caption{The decay schemes for several odd-even isotopes have
been illustrated for both the experimental measurements and the theoretical
calculations. We show the spin-parities, excitation
energies and log$ft$ values for several decay branches which have the largest branch
ratios. The parameters adopted here are $g_{pp}=1$ and $g_A(g_V)=0.5 g_{A0}(g_{V0})$.}
\begin{tabular}{|c|c|ccc|ccc|}
\hline
                   &                     &     & Exp.\cite{nucldata}& & & Theo.& \\
                   &  $J^{\pi}_i$ & $J^{\pi}_f$ & $E_{ex}$ & log$ft$ & $J^{\pi}_f$ & $E_{ex}$ & log$ft$ \\
\hline
                   & $7/2^+$ & $7/2^-$ & 0.051 & 6.88     & $7/2^-$ & 0        & 6.63 \\
$^{139}$Cs &                & $9/2^-$ & 1.283 & 7.4       & $9/2^-$ & 0.824 & 6.77 \\
                   &     & $5/2^-,7/2^-$ & 2.349 & 7.3       & $7/2^-$ & 3.308 & 6.66 \\
\hline
                   & $3/2^+$ & $3/2^-$ & 0.051 & 5.63     & $3/2^-$ & 0.333 & 5.18 \\
$^{203}$Au &                & $3/2^-$ & 0.225 & 6.61     & $1/2^-$ & 0        & 7.16 \\
                   &                & $5/2^-$ & 0        & 5.19     & $5/2^-$ & 0.418 & 5.11 \\
\hline
                   & $3/2^+$ & $1/2^-$ & 0        & 5.79     & $1/2^-$ & 0        & 7.06 \\
$^{205}$Au &                & $3/2^-$ & 0.468 & 6.43     & $3/2^-$ & 0.336 & 5.15 \\
                   &                & $5/2^-$ & 0.379 & 6.37     & $3/2^-$ & 0.761 & 5.07 \\
\hline
                   & $1/2^+$ & $1/2^-$ & 0.379 & 5.11     & $1/2^-$ & 0        & 5.02 \\
$^{207}$Tl\footnote{For this nucleus, we used the BCS coefficients extract from shell model to replace the solved bcs coefficients 
which are poorly reproduced near double magic nuclei.}
                   &                & $3/2^-$ & 0.898 & 6.16     & $3/2^-$ & 0.624 & 6.56\\
\hline
\end{tabular}
\label{ff128oe}
\end{table}


For the residual interactions, we adopt the Br\"uckner G-matrix derived from the
CD-Bonn potential~\cite{Mac89}. Two different parts of the interaction are
included, namely the particle-hole channel and particle-particle channel
interactions.
The two body matrix elements are calculated with Harmonic-Oscillator radial
wavefunctions.
Renormalization of the matrix elements has been introduced
to empirically take
into consideration the effect of the truncations of the model space over the
infinite Hilbert space into those orbitals in model space.
For simplicity, two overall factors are introduced,
namely $g_{ph}$ (for the particle-hole channel) and $g_{pp}$
(for the particle-particle channel).
The parameter
$g_{ph}$ mainly determines the position of the Giant Gamow-Teller resonance
(GTR).
Since there is no coupling between the GTR and the low-lying states the
GT beta decay to low-lying states is not affected by the value of $g_{ph}$.
Although for large $Q$ values some of the decay could go to the lower part of the GTR.
This  
contributes little to the total decay width due to their higher energies,
but it could be important for the beta delayed neutron decay branches.
A value of $g_{ph}$=1 which reproduces the experimental energy is adopted here.

The other parameter, $g_{pp}$, affects the energies of low-lying excited states
and
the GT matrix elements to these states. The $1^+$ excitation energies are
sensitive to
the value of $g_{pp}$. For nuclei near N=82, $g_{pp}$
values around $0.8$ best reproduce details energy levels (see Table \ref{cd}).
For nuclei near N=126 there is limited data on the final
state energies (especially for $1^+$), as we shall show later, with the quenching values we chose seems to agree with the experimental results, we find $g_{pp}=1.0$ better reproduces the half-lives. 
These parameters are obtained for the even-even nuclei, and
for consistency the same values are used in odd-mass nuclei.

In this work, we also introduce a quenching effect since our calculations show
an overall underestimation of half-lives compared with the experimental
measurements
for cases where there is good agreement for excitation energies.
This quenching of the QRPA calculations for the
spherical nuclei may have two origins.
The GT strengths obtained with shell-model calculations in the $sd$ shell~\cite{sdgt} and $pf$
shell~\cite{pfgt} are systematically larger than experiment by about a factor of two.
The ``quenching" of experiment relative to theory by about a factor of 0.5
is consistent with results obtained by second-order perturbation theory
corrections that take into account the excitation of nucleons outside of these model
spaces~\cite{arima,towner}.

Secondly, in spherical QRPA calculation, only one-phonon excitations have been
taken into account, while shell-model calculations show that there exists the mixing
between the single- and multi-phonon states. 
Two kinds of mixing outside of our model could be present: 
the low-lying part mixes with the GTR which shifts strength to higher energy; 
and the charge conserving 1p-1h excitations coupled with charge exchange 
GT excitations which spreads the GT strength and shifts some of the strength to higher energies.
We determine the quenching empirically by 
comparing the calculated log$ft$ values and
half-lives with the experimental
ones for the Cd isotopes. The results are shown in
table \ref{cd}. The log$ft$ values are for
individual final states, while the half-lives take into account the
decay to all final states (GT is the dominate decay mode).
A choice of
$g_A/g_{A0}\approx0.4$ generally reproduces the half-lives.
The largest difference comes from $^{130}$Cd due to the fact that
we predict a smaller excitation energy compared to experiment.
From this comparison, we adopt the value of $g_{A}/g_{A0}=0.4$ for the GT
calculation as an optimal choice in our calculation for both $N=82$ and $N=126$ regions.

Compared to the simple operator $\sigma$ operator for GT decays, the FF operators  
are more complicated with several different spin-parity combinations. 
As was shown in \cite{War91}, the tensor part of the $0^+$ to $0^-$ FF operator 
is enhanced due to mesonic-exchange currents. As in  Ref. ~\cite{ZCCLMS13}
we use an enhancement factor$\epsilon=2$. 
In general, 
we could use different quenchings for the other types of FF operators as was
done for the shell-model calculations of Ref. ~\cite{ZCCLMS13}.
However, this is quite complicated and even with least-square fits in Table. I of Ref. ~\cite{ZCCLMS13}, 
we still see some large deviations,
 We cannot simply use the same quenching adopted in the shell-model
approach since the origins of quenching for the shell-model and QRPA are different. 
For the shell model it comes from the 
model-space truncation (with a rather small model space compared with QRPA calculation), while for QRPA, it is from the configuration-space truncation (only one-phonon excitations have been taken into account in QRPA). 
In this sense, different queching schemes should be used. For simplicity 
we use an overall quenching factor for all the FF operator and we vary 
this value to find the best agreement to a rather limited set of data.

The quenching factors $g_A/g_{A0}=0.5$ and $g_V/g_{V0}=0.5$ 
for FF transitions are obtained this way by comparing 
the theory to experiment for some relatively strong FF transitions
to low-lying states. In Table \ref{ff82ee}, 
the comparison between experiment
and theory with and without the chosen quenching factors for the 
decay of Te and Xe isotopes is
shown. Without quenching, there is a general underestimation about 
$0.6$ in the log$ft$ values. When we use above queching, for most decay branches, 
this deviation reduces to less than $0.3$. The same quenching factors are used in 
the N=126 region. 


For the odd-mass nuclei, the same parameter sets ($g_{pp}$ and $g_A$) are
adopted. The results for some FF branch ratios are shown in Table \ref{ff128oe}. 
The deviation of the log$ft$ values 
are generally larger than the even-even case. We see reasonable agreement for 
$^{139}$Cs and $^{203}$Au. 
The  $^{207}$Tl to $^{207}$Pb decay is particularly simple.
When experimental single-particle energies are used 
the transition is dominated by the $3s_{1/2}$ to $2p_{1/2}$ contribution
and we obtain log$ft$=5.11 (with quenching) compared to the
experimental value of 5.02. 
In general, our QRPA approach is not very good for cases with one or two nucleons removed
from the double-magic nuclei where the pairing is weak and the BCS solution 
sometimes overestimates the pairing effect. Also
transitions to a few low-lying states
are sensitive to the precise value of the single-particle energies.
(The SkX single-particle
energies differ from experiment by up to 0.5 MeV, see Fig. 4 in \cite{SKX}.)

For the global parameters such as Q values and
neutron separation energies, we use the experimental values if they are available,
otherwise we use the masses predicted by some phenomenological mass models. For
comparisons, we used two mass model here, the FRDM model~\cite{MPK03} and
HFB21~\cite{GCP09}.

\section{Results and Discussion}

\begin{figure*}
\includegraphics[scale=0.4]{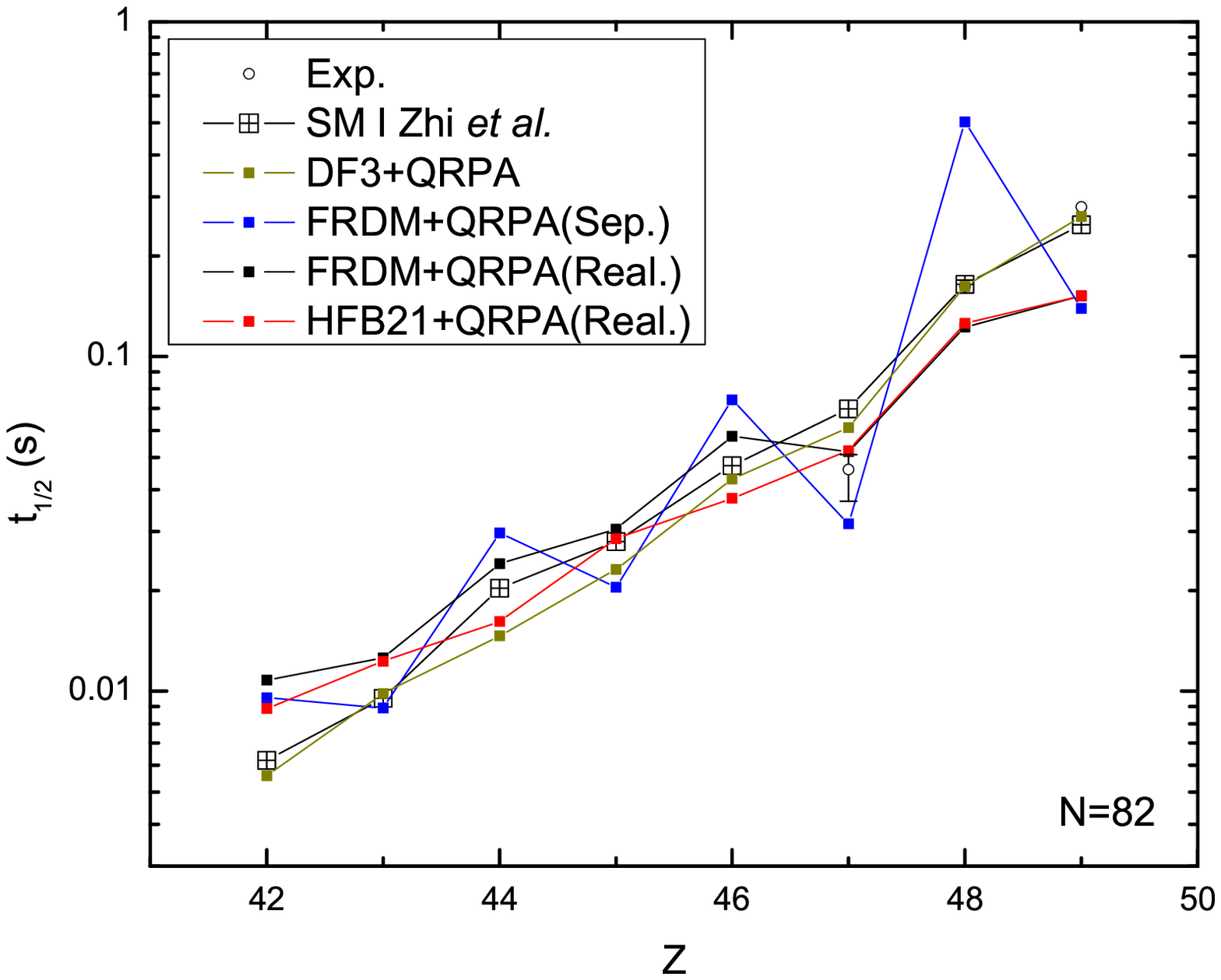}
\includegraphics[scale=0.4]{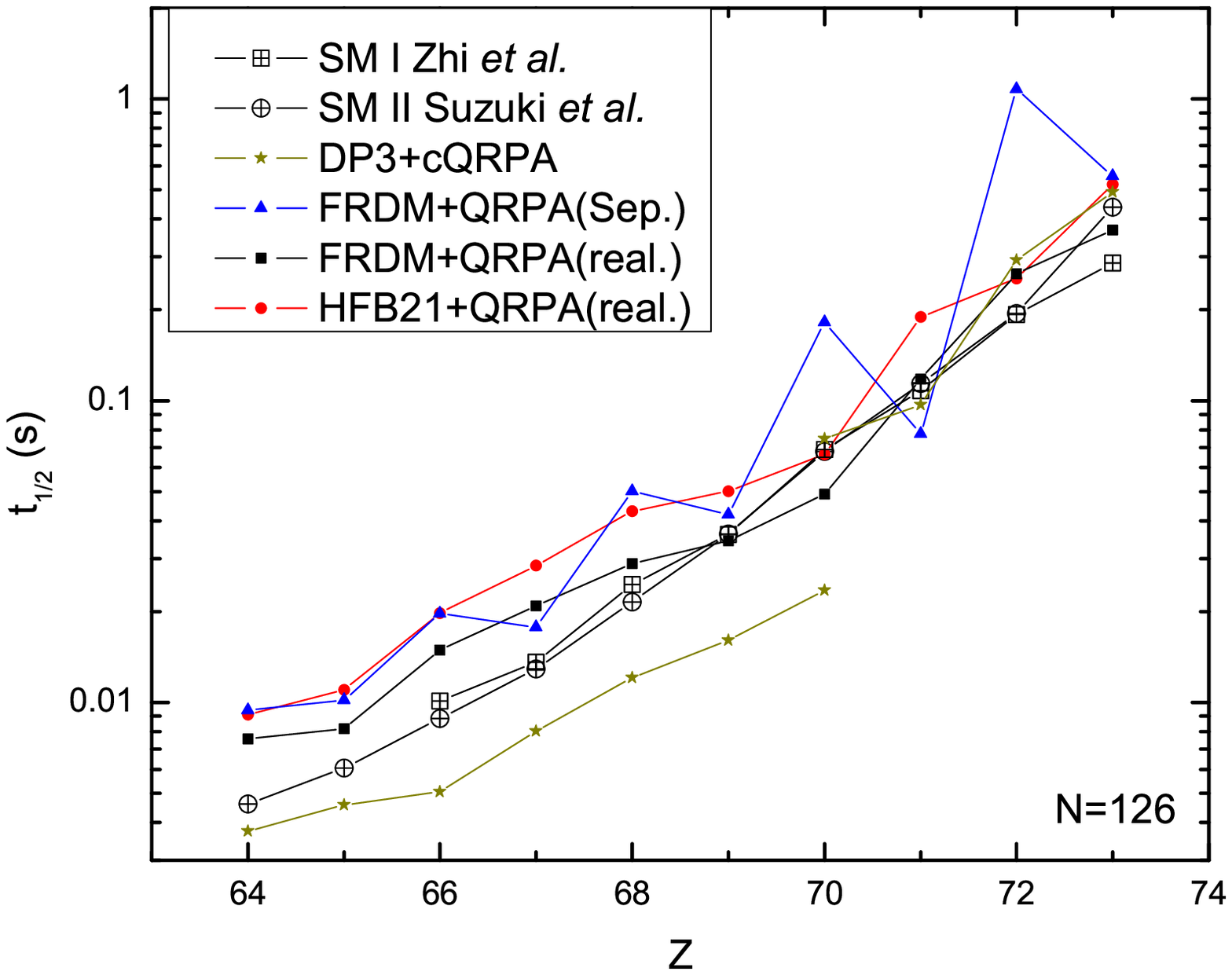}
\caption{(Color online) Comparison of half-lives between shell-model calculations:
SMI~\cite{ZCCLMS13}, SMII~\cite{SYKO12} and QRPA calculations: with separable
forces in~\cite{MPK03}, continuum QRPA in~\cite{Bor03,Bor11} and this work for
N=82 and N=126 isotonic chains.}
\label{tcmp}
\end{figure*}

\begin{figure*}
\includegraphics[scale=0.4]{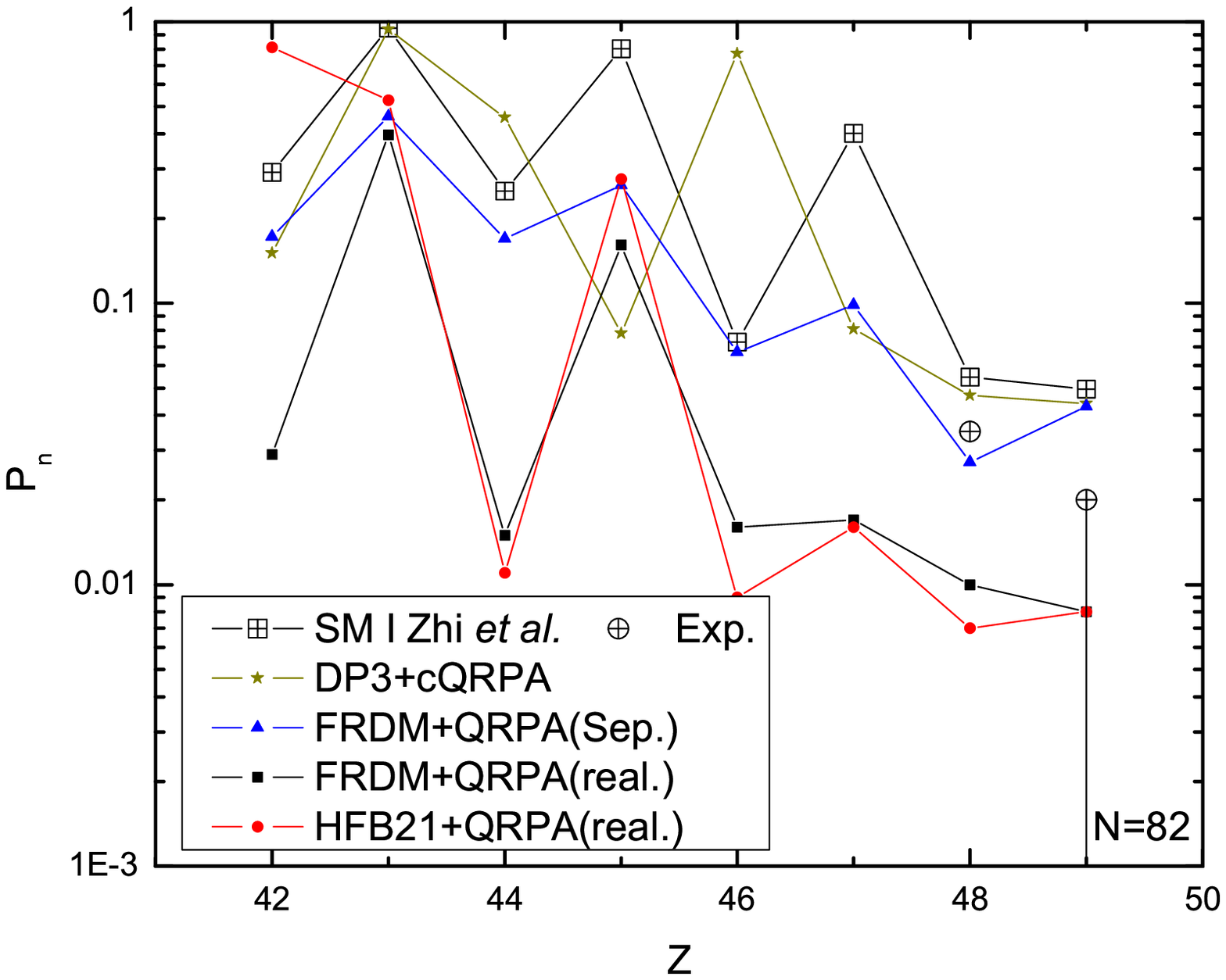}
\includegraphics[scale=0.4]{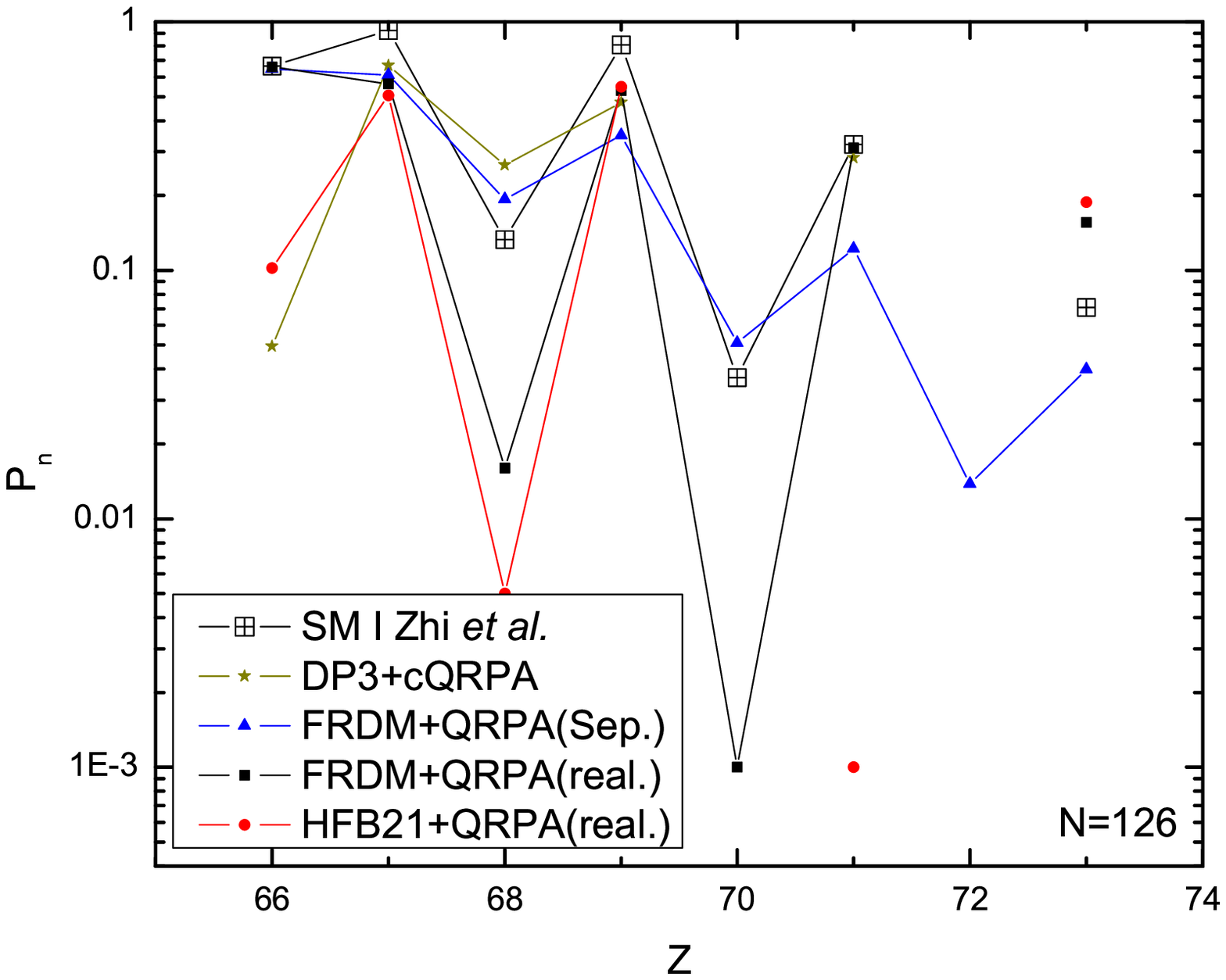}
\caption{(Color online) Comparison of total $\beta$-delayed neutron emission probability
between shell-model calculations~\cite{ZCCLMS13,SYKO12} and QRPA calculations
in~\cite{MPK03,Bor03,Bor11} and this work for N=82 and N=126 isotonic chains.}
\label{pncmp}
\end{figure*}

\begin{figure*}
\includegraphics[scale=0.4]{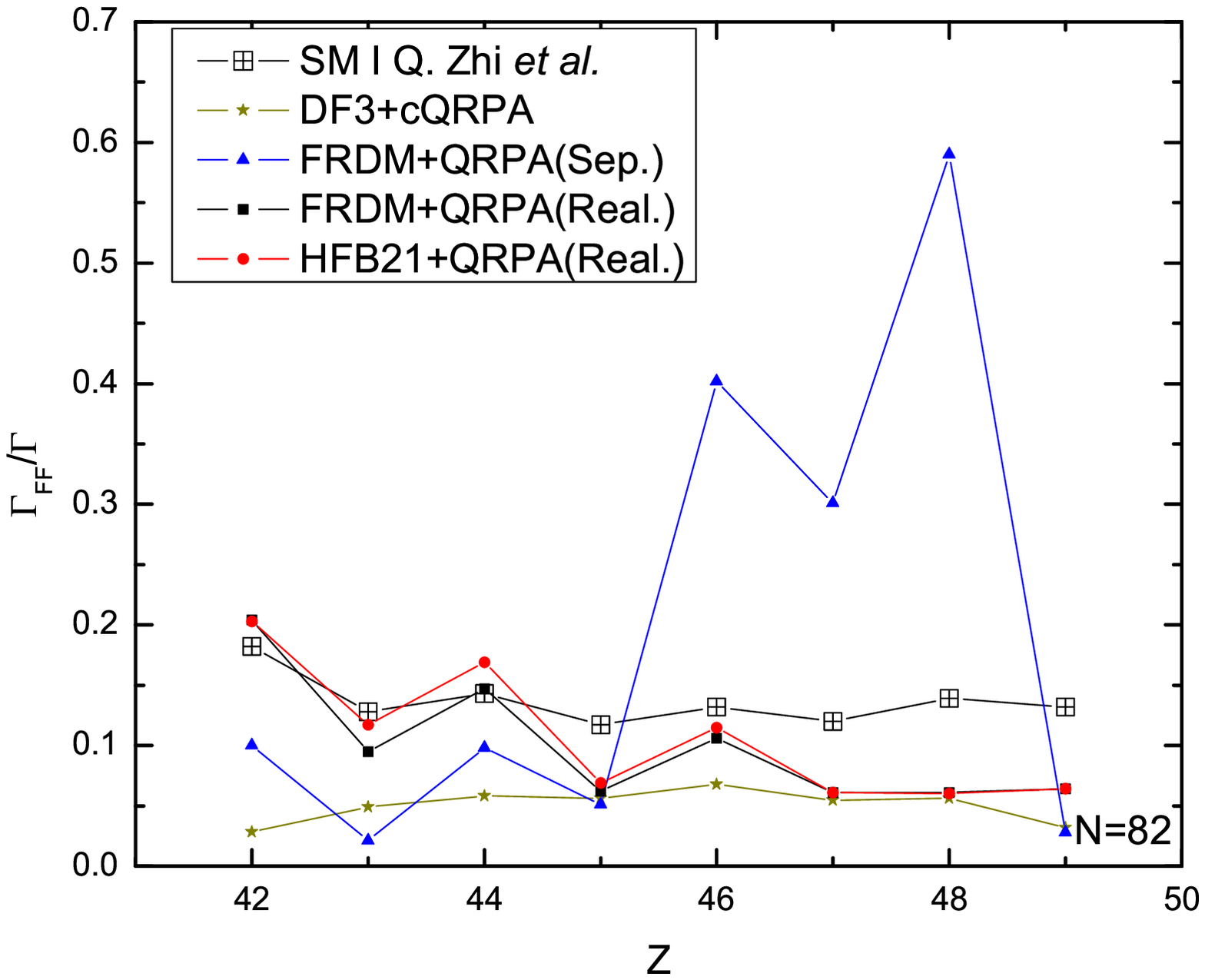}
\includegraphics[scale=0.4]{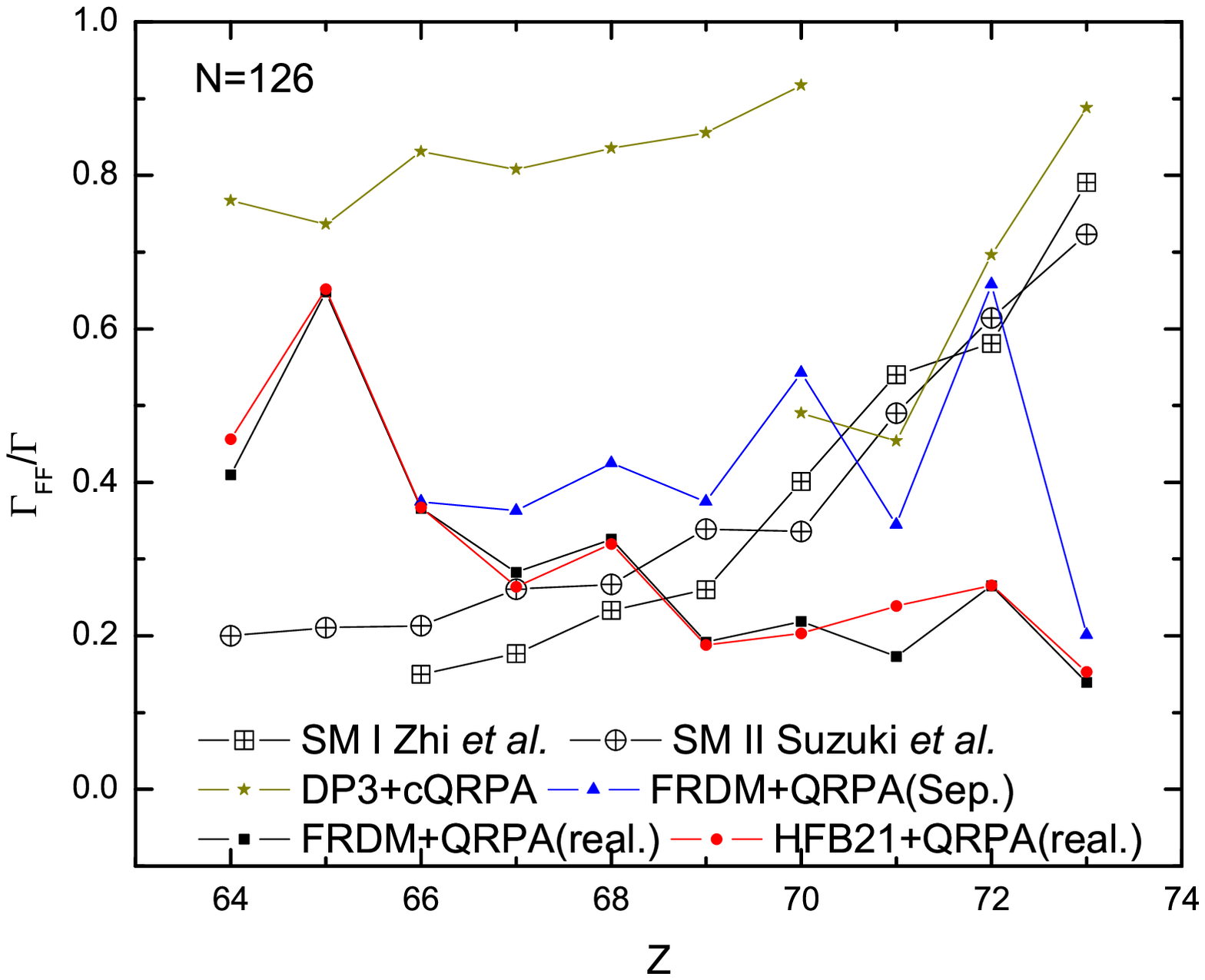}
\caption{(Color online) Comparison of ratios of first-forbidden part of the total decay width
between shell-model calculations~\cite{ZCCLMS13,SYKO12} and QRPA calculations
in~\cite{MPK03,Bor03,Bor11} and this work for N=82 and N=126 isotonic chains.}
\label{ffcmp}
\end{figure*}

With the determined parameters, we proceed to the calculations of isotonic
chains
on the r-process path in the vicinity of $N=82$ and $N=126$.
We first
show the results with neutron magic number $N=82$ and $N=126$ where the shell
model results for these isotones are available. In
Figs. \ref{tcmp}-\ref{ffcmp}, we show the comparison of our
results with shell-model calculations from Ref.~\cite{SYKO12,ZCCLMS13} (SMII results\cite{SYKO12} are 
updated by making a correction of a calculations error) and other
QRPA calculations such as the QRPA with separable force from Ref.~\cite{MPK03}
and the
continuum QRPA from Ref.~\cite{Bor03,Bor11}. For the first-forbidden (FF) part
of the
decay, Ref.~\cite{MPK03} uses the gross theory instead of microscopic
calculations, while for all the other calculations, the FF parts are calculated
explicitly. 

We first compare the most important observables -- the half-lives
predicted by different methods in Fig. \ref{tcmp}. For $N=82$ isotones, there is good
agreement among different methods except those from separable-force model
of Ref.~\cite{MPK03} for which
there are systematic overestimation of half-lives on the even-even isotones
compared with other methods for the high Z (low Q) isotones. This is due to their over-estimation of 
the excitation energies and underestimation of the matrix elements due to 
the lack of particle-particle interactions for QRPA. This results in an odd-even 
staggering behaviors for the half-lives
in separable-force calculations. However, when the
Q becomes large, this effect is reduced as we will see later. Except for Ref.~\cite{MPK03}, the
general discrepancy among different methods for most nuclei is within a factor
of two. This shows a good convergence of the predicted half-lives in
microscopic calculations. While comparing with the experiments, our
method underestimates the half-lives for $^{131}$In and $^{130}$Cd, especially
for the former. The reason for this is that the BCS method does not
work well near the doubly-magic nuclei where the pairing is weak and the BCS solution 
sometimes overestimates
the pairing effect as we mentioned previously. For $^{130}$Cd the reason for the disagreement
is due to the energy of $1^{+}$ state being too low compared to 
experiment as seen in Table \ref{cd}. However, we find better
agreement for $^{129}$Ag, compared with other methods. The results with FRDM
mass model seem agree well with the shell-model calculations while those of the HFB21
mass give a smoother behavior over the isotonic chains. Compared with the shell-model calculation, 
we find the trend that when Q value becomes larger, our results 
give some overestimation on the half-lives.

For $N=126$ isotones shown in Fig. \ref{tcmp}
the discrepancy between different methods becomes larger. Again, there is a
systematic overestimation for results from the separable-force model Ref.~\cite{MPK03} for even-even nuclei for larger Z.
For the continuum-QRPA method, two sets of results are shown; 
the results from ~\cite{Bor03} systematically underestimate the
half-lives, while the results from Ref.~\cite{Bor11} are closer to other calculations.
The two Shell Model calculations give similar results for the half-lives. The difference for the two shell-model calculations comes from the different
quenching factors and model space they adopted that produces a nearly small constant difference 
within a factor of about two. In our calculations, there is difference of less than a 
factor of two with the two different mass models. The results with the 
FRDM masses shows a good
agreement with the shell-model calculation while those with HFB21 masses
give longer half-lives. For most of the nuclei here, with the FRDM mass model, our results 
differs with the shell model by a factor less than two. Overall, we find good agreement for the half-lives
for the $N=126$ isotones among the different methods except those from
Ref.~\cite{MPK03} at high Z and Ref.~\cite{Bor03} at low Z. 

Next we compare the results for the 
$\beta$-delayed neutron emission probability $P_n$ in Fig. \ref{pncmp}. For the $N=82$ isotones,
unlike the half-life results, there are now larger differences among different
methods. There is a general trend that the odd-even nuclei have larger $P_n$
values than their neighboring even-even nuclei except results from
Ref.~\cite{Bor03}. We predict lower $P_n$ values than the other
calculations, especially for the even-even nuclei. For $^{130}$Cd, our
result is smaller than the experimental value of 3.5\%.
In this case the neutrons come from beta decay to the region of
excitation in $^{130}$In below the beta-decay Q value of 8.9 MeV and above
the neutron decay separation energy of 5.1 MeV. Our low $P_n$
value indicates that
the QRPA GT distribution in the region of 8 MeV 
is not spread out enough due to the lack of
coupling with particle-hole states.
The FRDM+QRPA(Sep.)~\cite{MPK03} seems to have better
predictions for the $P_n$ values, but this comes from the fact that
the energy of the lowest $1^+$ state is too high in this case as they predict a much longer half-life for this nucleus, that is, about four times larger than the experimental half-life. 

The same situation happens for the $N=126$ isotones (right panel of Fig. \ref{pncmp}). The differences
among the calculations are quite large for even-even isotones, while there
is better agreement for the odd-even cases.
In the region $Z<70$, the $P_n$ values are
close to $1$. This can be important for the r-process evolution. 
The shell-model calculation~\cite{ZCCLMS13} predicts larger $P_n$ values than most other
calculations for both $N=82$ and $N=126$ cases, the reason for this is that in shell-model calculations, due to
their large configuration space, the strength is fragmented compared with QRPA
calculations and more strength has been distributed above the neutron separation
energy threshold.

In order to see the importance of inclusion of the first-forbidden part, we compared
the ratios of FF part to the overall decay width in Fig. \ref{ffcmp}. For $N=82$, except
Ref.~\cite{MPK03} (as they haven't calculated the FF parts explicitly), the three methods (QRPA, cQRPA and shell model) predict generally 
similar ratios. We have lower FF ratios which are close to those from cQRPA for high Z 
and higher FF ratios which are close to shell-model calculations for low Z nuclei. 
On the other hand, this ratio is about the same for the two mass models in our calculation
There is a systematic difference between the cQRPA and shell-model calculations 
and our results are in between. In general, for the $N=82$
isotonic chain, all the calculations show that the first-forbidden part plays a less
important role in $\beta$-decay. However, this is not the case for the $N=126$ 
(right panel of Fig. \ref{ffcmp}) isotonic chain, where there are large discrepancies among different methods.
We obtain FF ratios that decrease with Z in contrast to the other methods where they increase.  
Comparing our results carefully with those of Fig. 16 in Ref.\cite{ZCCLMS13}, 
we found the difference comes from the fact that the two methods have different GT strength distributions. 
Although the excitation energies for $1^+$ states are similar 
(around 2 MeV for both methods for three nuclei $^{194}$Er,$^{196}$Yb and $^{198}$Hf), the structures of these states are different. In QRPA calculation, we see an 
small increase of the energies for the first $1^+$ states as the proton 
number increase, but the energies of the $1^+$ states which give 
dominant contribution to the GT decay however decrease sharply with the 
increasing proton number. This is not seen in shell-model calculations. 
This means that there may be too much strength presented in the 
low-lying states for N=126 isotones with higher Z due to the 
configuration space truncations. But on the other hand, shell-model 
calculations have too limited model space and could 
result in this different behaviors of GT strength distributions as well. 
Although Ref.~\cite{Bor03} has also a very large FF ratio, compared with the
same method in Ref.~\cite{Bor11}, this would be just due to a wrongly accounting
of parameters in their calculations and their FF ratios for low Z isotones may need some updates. 
On the other hand we see that in the two
shell-model calculations, the different quenchings and model spaces also give some different ratios as they
have similar quenching for the GT part. So as explained above, the difference
between our calculation and the shell-model calculations may came from two
aspects, the different configuration space and the different model space, and
further investigation is needed to explain the discrepancy for the ratios of FF
parts.

\begin{figure*}
\includegraphics[scale=0.4]{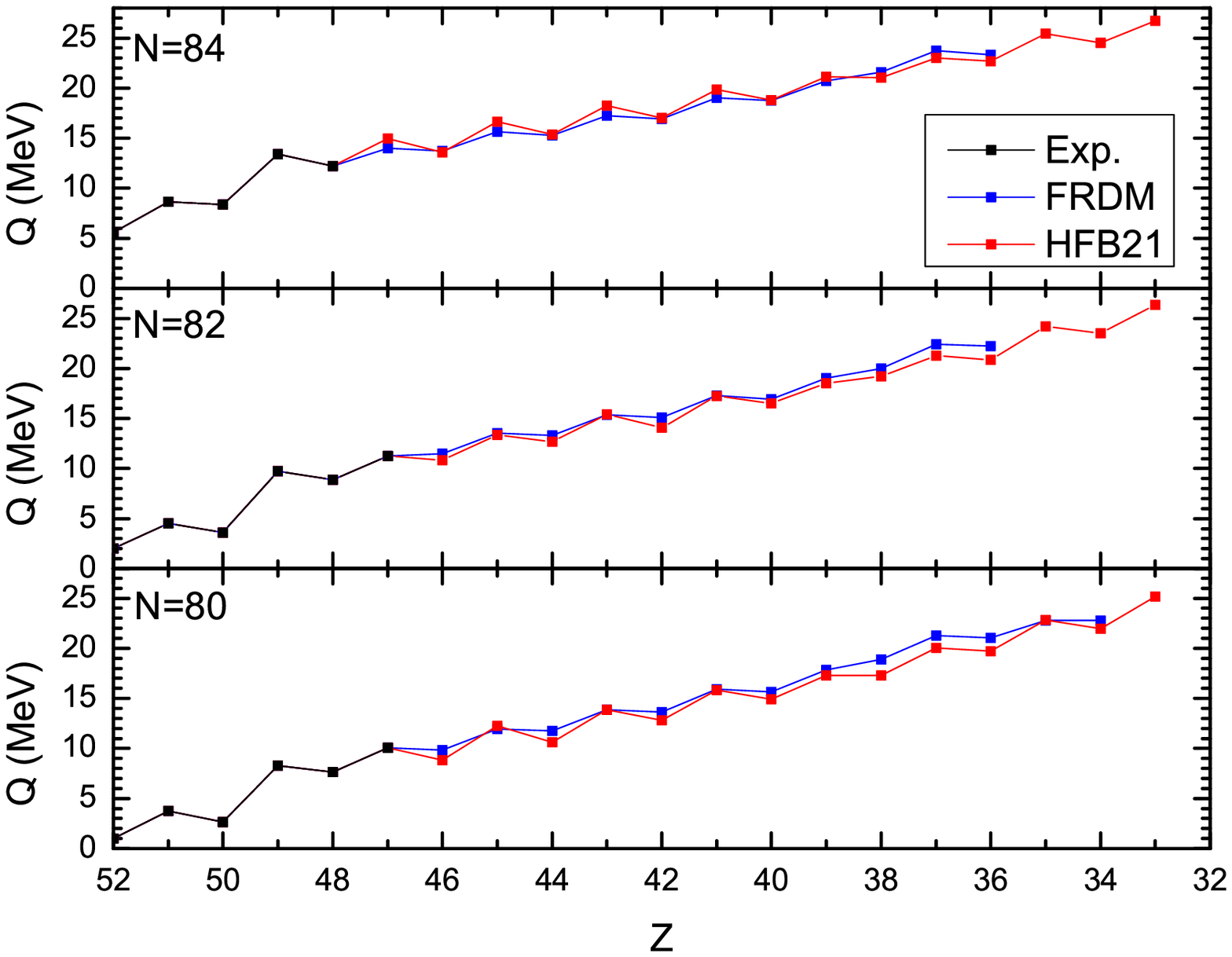}
\includegraphics[scale=0.4]{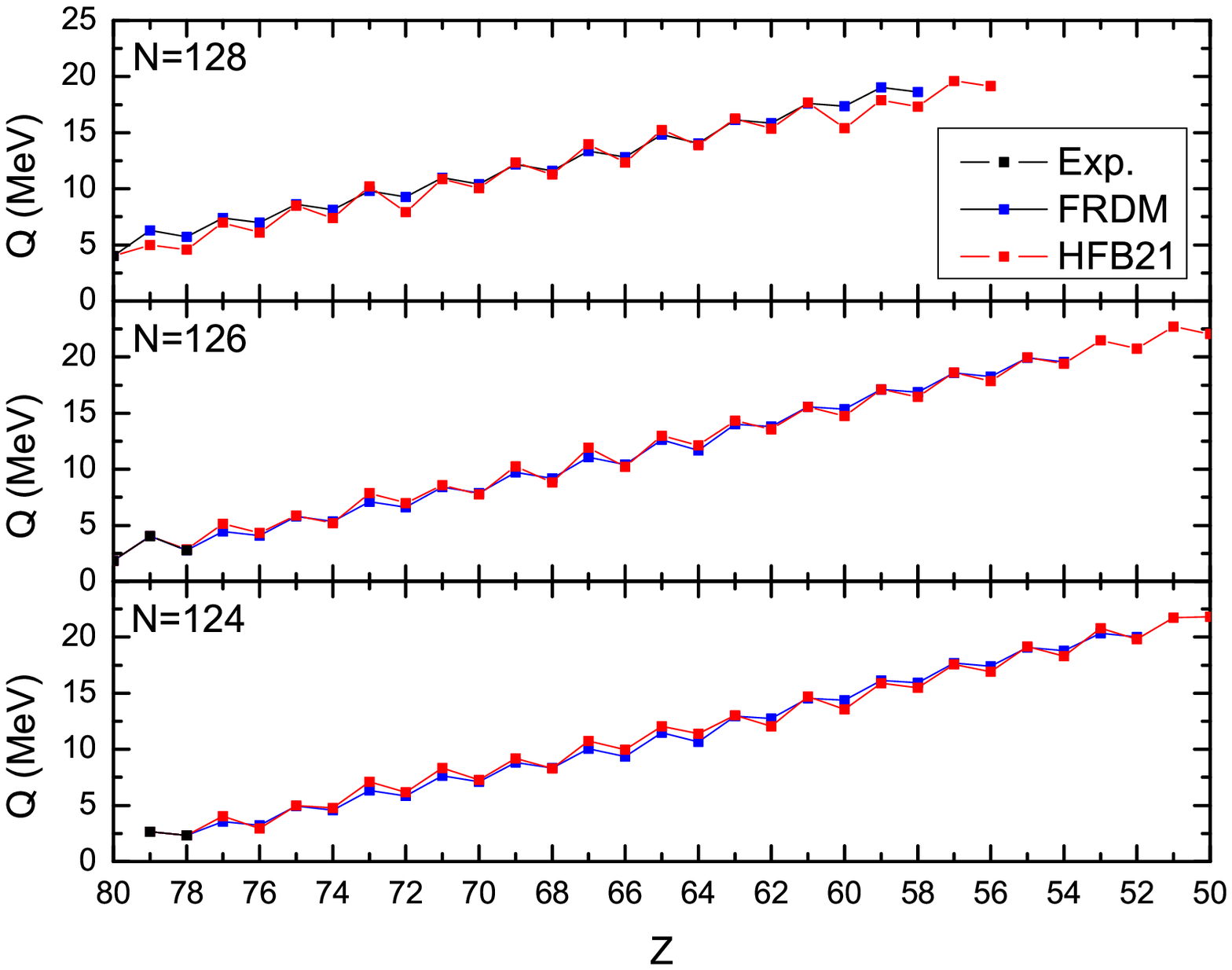}
\caption{(Color online) The Q values we adopt in our calculation for $N=80,82,84$ and $N=124,126,128$ isotones, the meanings of the symbols are explained in the text.}
\label{Q82}
\end{figure*}

\begin{figure*}
\includegraphics[scale=0.4]{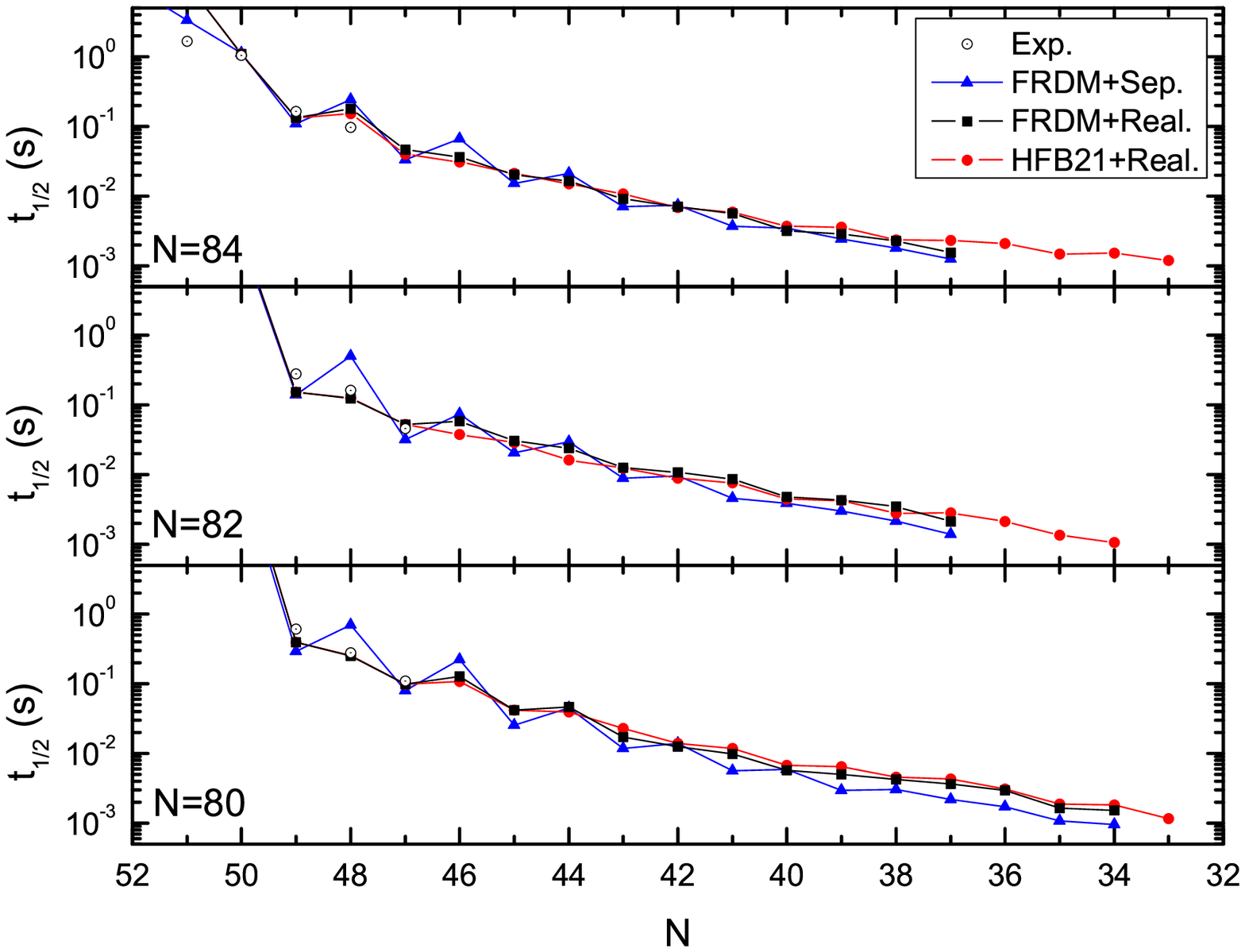}
\includegraphics[scale=0.4]{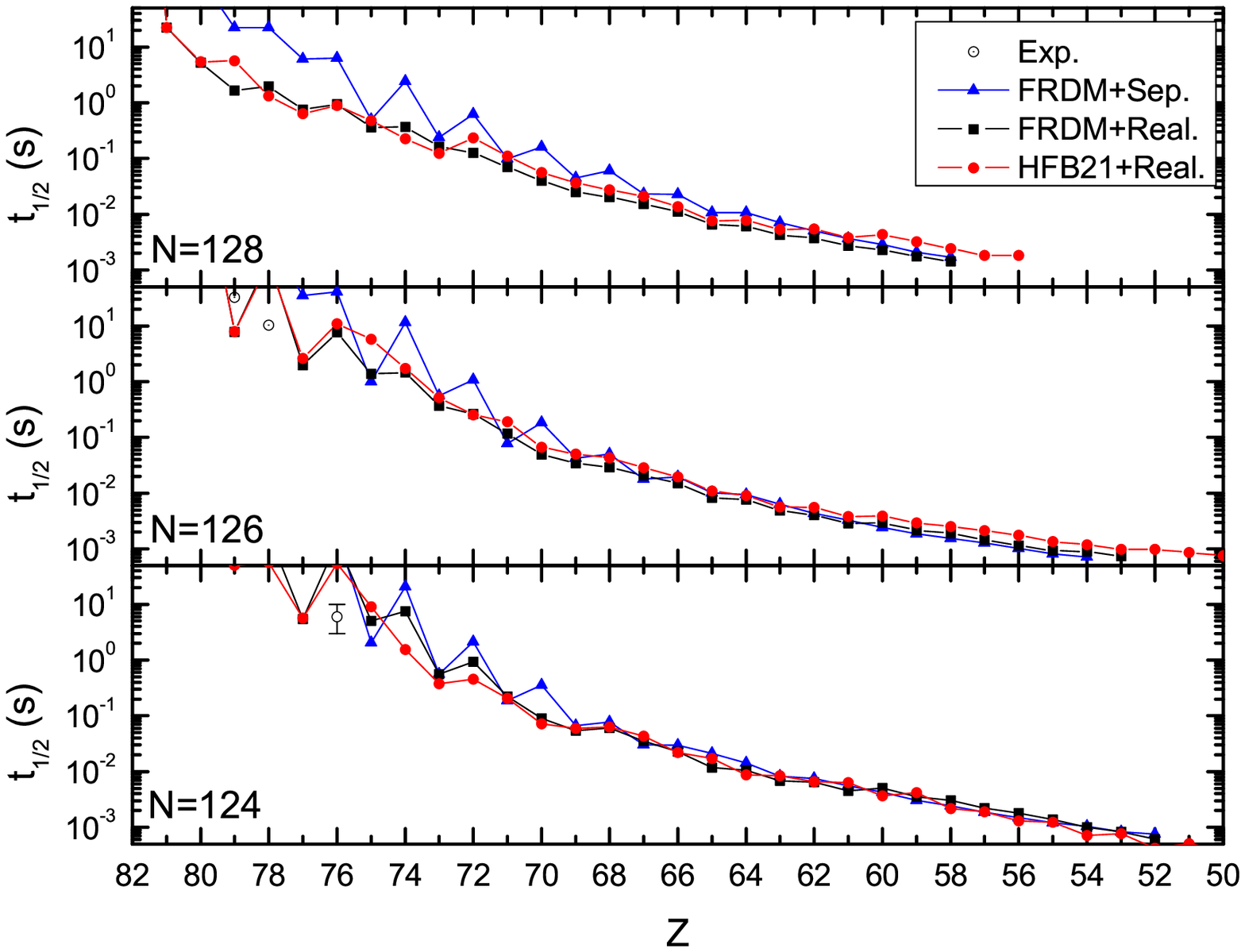}
\caption{(Color online) The calculated half-lives from Ref.~\cite{MPK03} and our calculations
with two mass sets for N=80,82,84 and N=124,126,128 isotones. The comparison
between experiments and theories has been presented.}
\label{t82v}
\end{figure*}

The advantage of QRPA methods is that we can calculate with a larger model space
more nuclei than those on the magic number line as in the shell
model calculations. So we present more results around the magic number vicinity
for even-end nuclei. First, we present the Q value sets we use in our
calculation in Fig. \ref{Q82}. As mentioned above, we adopt the Q values if they are available
from experimental measurements as those in ~\cite{AWT02}. Otherwise, we use two
mass models for the sake of comparison, the FRDM~\cite{MPK03} mass model come from the macroscopic droplet
models and the HFB21~\cite{GCP09} model from self-consistent microscopic
Hartree-Fock-Boglyubov calculations. From Fig. \ref{Q82}, we see that around the neutron number
$N=80$, the Q-values obtained from the two mass tables are basically the same for
most even-even nuclei while there are $\sim 1MeV$ differences for most odd-even
ones, generally FRDM predicts smaller Q-Values except for those nuclei with lower Z.
However, for $N=82$ and $N=84$, the contrary happens and FRDM generally predicts
larger Q-values. As for $N=124$, the same situation happens as for $N=82$, large
difference occurs at low Z. As For $N=126$ and $128$, the discrepancy from the two
mass models becomes irregular.

\begin{figure*}
\includegraphics[scale=0.4]{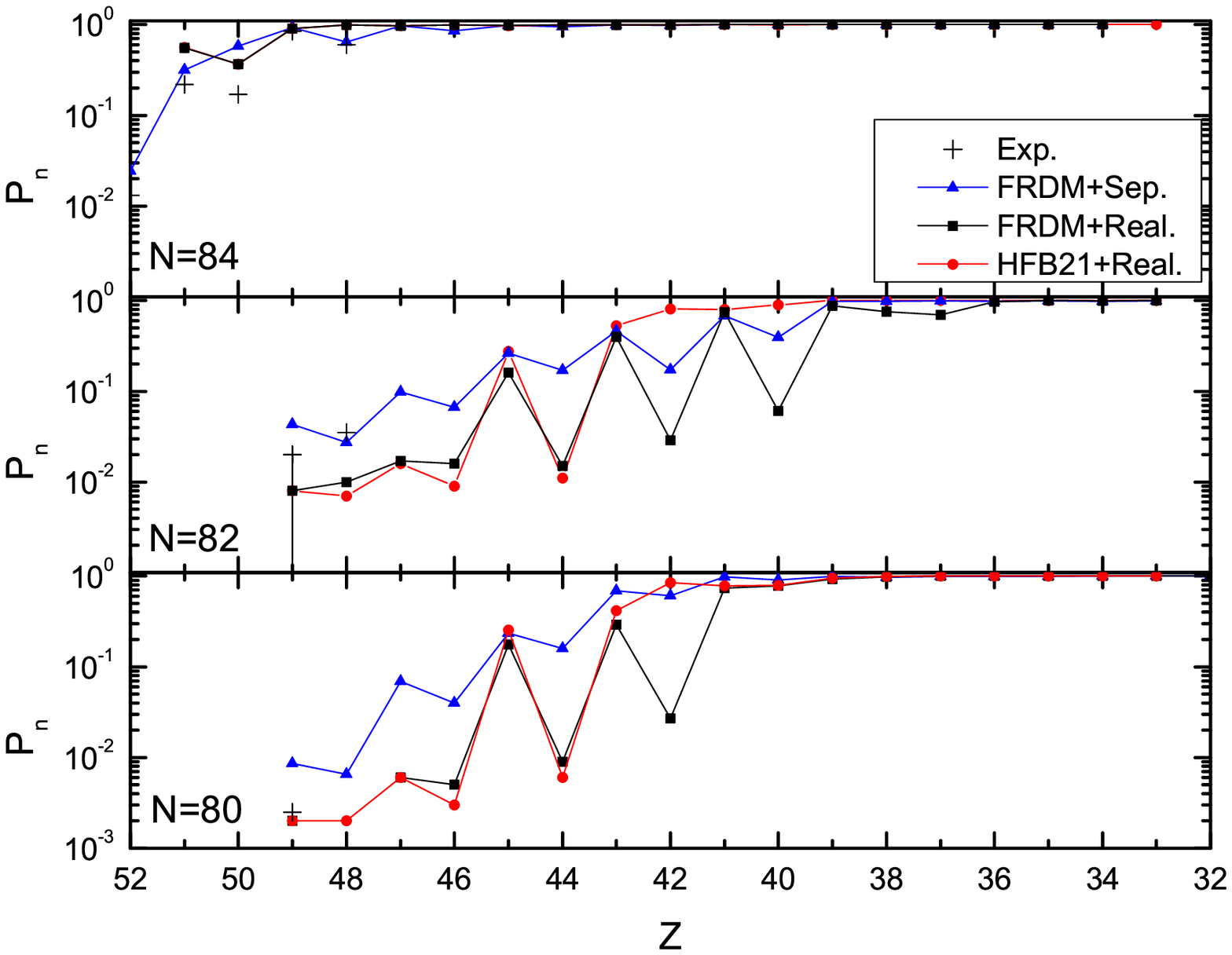}
\includegraphics[scale=0.4]{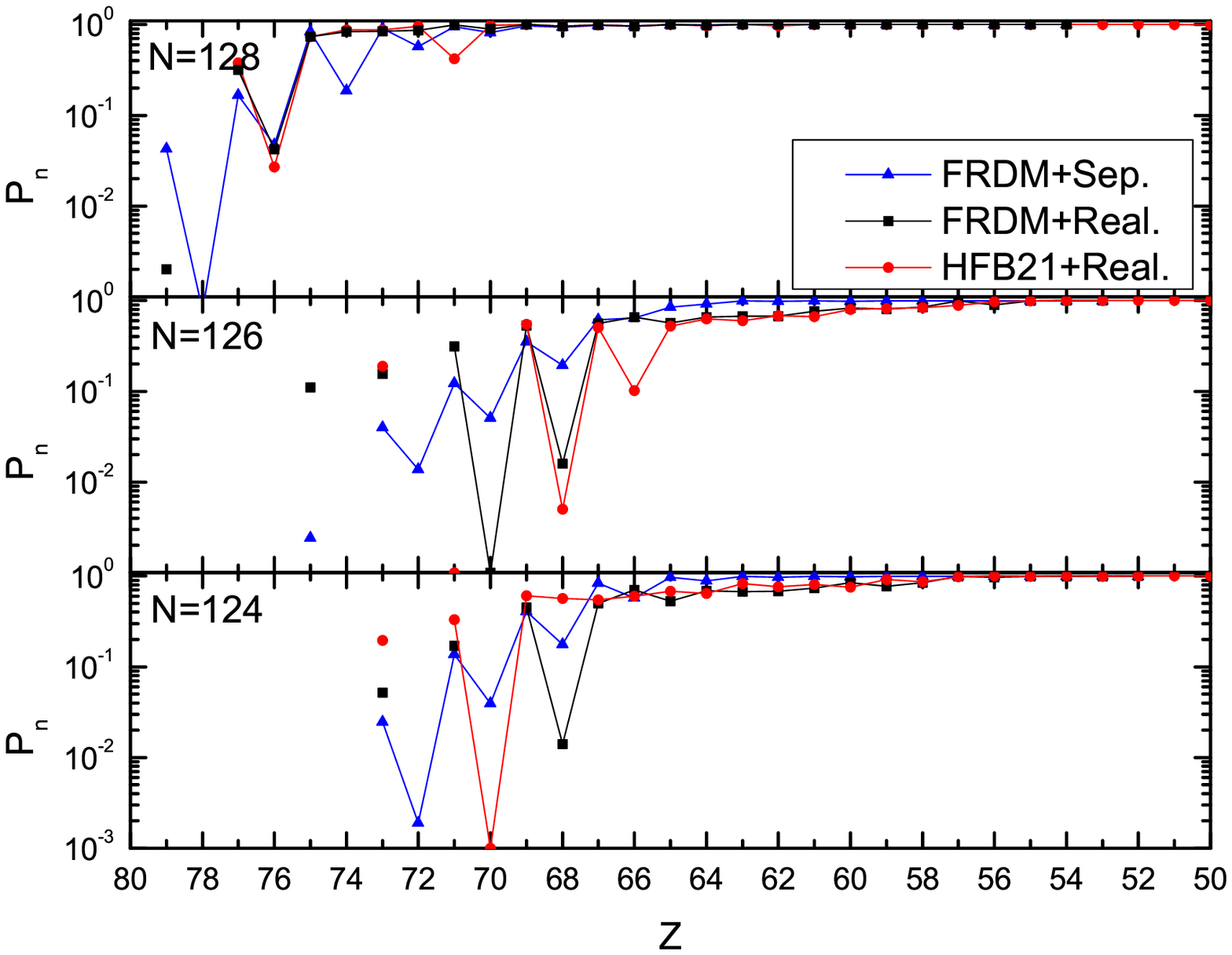}
\caption{(Color online) The calculated $P_n$ values from Ref.~\cite{MPK03} and our calculations
with two mass sets for N=80,82,84 isotones. The comparison between experiments
and theories has also been presented.}
\label{pn82v}
\end{figure*}

We calculated the half-lives of the isotones with the two mass models and the
results are presented in Fig. \ref{t82v}.  For $N=$80, 82, 84 isotones, we make
comparison with those results in Ref.~\cite{MPK03} as well as some experimental values. When compared with the
experimental results, we find very good agreement. In general, the agreement
gets better when the nuclei are far away from the proton magic number $Z=50$.
The results with $N=80$ and 84 has the same trends as N=82 we analyzed
previously. Compared with results obtained in Ref.~\cite{MPK03}, we would see
the
same behaviour of overestimation of even-even nuclei in their calculation with
moderate Q values. When the Q value increases the results in both calculations
get closer and for much larger Q values, Ref.\cite{MPK03} predicts shorter half-lives than ours. 
The reason for this is due to the different treatments of the excitation energies. In
Ref.~\cite{MPK03}, QRPA energies have been used directly as the excitation
energies, which are in fact about several MeV higher than the actual excitation
energies. Although the strength spreading has been introduced, even without the
quenching effect, the deviation is still large. If Q values are large, then the effect from this difference between QRPA and excitation energies is
weakened. Meanwhile the quenching reduces our decay strength and make our predicted half-lives longer than theirs. While in the case of odd nuclei, the single particle transition may
play dominant role for low-lying states, we would not see this large deviation of half-lives for isotones with large proton numbers (Z).
Generally, for
low Q values, deviation occurs for even-even nuclei and agreement is achieved
for odd-even nuclei, and for large Q values, we have longer half-lives due to the inclusion of quenching. If we compare the results obtained for the two mass models,
we see that for the same nuclei, large Q value shortens the half-life.  However,
exceptions do exist as the mass model also affect the empirical gap parameters
and hence change the structure indirectly. Such as the case of $^{132}$Cd, with the
same Q value we have a small difference with two mass sets due to the different
gap parameters we obtained. Different mass sets produce different half-lives and
their impact on r-process is  to be investigated. Also, the smoothing behavior
of the half-lives on the proton number Z depends on the mass set one chooses.
A choice of mass sets may produce the unwanted even-odd staggering behavior. It is
somehow hard to tell, whether the staggering comes from the microscopic
approaches or the Q value sets, due to the uncertainty of the theory.
The same conclusion can be drawn on $N=124$, 126, 128. One has only limited
experimental data in this region, and the agreement for these limited data with
the calculations are very poor, as the Q values for these nuclei are very small.
For $N=$124 and 126, with the increasing Q values, our calculation tends to agree
with calculations of Ref.~\cite{MPK03} for odd mass nuclei, but for N=128, the
deviations always exist. In general, the different mass sets produce a deviation
less than a factor of two close to the deviations produced from different
methods we discussed for N=126 isotonic chain.

As for another measurable $P_n$, we have even less experimental data for comparison. Our calculation has smaller $P_n$ values compared with experiments, but the deviation is not large
especially when $P_n$ is close to $1$. For $N=80$ and 82, the $P_n$ values
increase as the decrease of proton numbers. For very neutron-rich nuclei, the
beta-decay is followed immediately with neutron emissions. We see an odd-even staggering
behavior as previously found in shell-model calculations. The deviation between our results
and those from Ref~\cite{MPK03} is large in magnitude but keeps the same trends
(the reason of this has been explained above as the strength has been shifted
to high-energies region systematically in their calculations). However, the two sets
of results seem to agree with each other when $P_n\sim 1$. This is because the
neutron separation energies here are pretty smaller, nearly all the strength
lies in the window between $E_n$ and Q. When one crosses the magic number line $N=82$,
there appear increased $P_n$ values as the nuclei here become less stable against
the neutron emission, which agrees with results in Ref.~\cite{MPK03}. For the
heavier nuclei region with N around 126, the neutron emission probability hasn't
been measured for any nuclei. The general trends here for the two calculations
again keeps the same while they differ in numbers as above. The nuclei beyond the
neutron magic number $N=126$ show the large $P_n$ values as for the region of $N\sim
82$.

Due to the drawbacks of the QRPA methods, limited accuracy has been obtained
especially for $P_n$ values. But one could see some improvements of the QRPA calculations, the half-lives are closer to the experimental values as well as those by shell model calculations compared with results in\cite{MPK03}.
For regions where shell-model calculations are absent, the decent results have
been obtained. To further improve the predicted $P_n$ values, the strength
spreading from multi-phonon effect could be introduced such as those from the
particle-vibration couplings~\cite{LR06,CSB10}.

\section{conclusion}
In this work, we calculated the weak decay properties of even proton number isotones near the neutron
magic numbers 82 and 126 with the spherical QRPA method with realistic forces.
Our results agree well with other calculations on the neutron magic number
chains 82 and 126. We give the predictions for more nuclei in these region and
compared with results in Ref.~\cite{MPK03}. Different mass models have been used for
the sake of comparison, and they produce a moderate difference for the final
results. In general, we make some improvements on the accuracy of decay rates compared with Ref.~\cite{MPK03} and
its impact on the r-process simulations are to be investigated.

This work was supported by the US NSF [PHY-0822648 and PHY-1068217].

\end{document}